\documentclass[12pt]{spieman}  % 12pt font required by SPIE;
\usepackage{amsmath,amsfonts,amssymb}
\usepackage{graphicx}
\usepackage{setspace}
\usepackage{tocloft}
\usepackage[colorlinks=true, allcolors=blue]{hyperref}
\usepackage[version=3]{mhchem}
\usepackage[normalem]{ulem}
\usepackage{multirow}

\usepackage{lineno}
 
\title{Mechanical strength and millimeter-wave transmittance spectrum of stacked sapphire plates bonded by sodium silicate solution}

\author[a]{Takayuki Toda}
\author[a,b]{Yuki Sakurai*}
\author[a]{Hirokazu Ishino}
\author[b]{Tomotake Matsumura}
\author[a]{Kunimoto Komatsu}
\author[b]{Nobuhiko Katayama}
\affil[a]{Department of Physics, Okayama University, Okayama, Japan}
\affil[b]{Kavli Institute for the Physics and Mathematics of the Universe (WPI), The University of Tokyo Institutes for Advanced Study, The University of Tokyo, Chiba, Japan}

\cftpagenumbersoff{figure}
\cftpagenumbersoff{table} 
\begin{document} 
%\linenumbers
\maketitle

\begin{abstract}
The polarization modulator unit for the low-frequency telescope in LiteBIRD employs an achromatic half-wave plate (AHWP). 
It consists of five layers of a-cut sapphire plate, which are stacked based on a Pancharatnam recipe. In this way, the retardance of the AHWP is a half-wave over a bandwidth of 34-161~GHz. 
The diameter of a single sapphire plate is about 500~mm and the thickness is 5~mm. When a large diameter AHWP is used for a space mission, it is important for the AHWP to survive launch vibration. 
A preliminary study indicates that the five-layer stacked HWP has a risk of breakage at the launch unless the five layers are glued together and mechanically treated as one disk. 
We report our investigation using a sodium silicate solution that can bond the sapphire plates. 
This technique has been previously investigated as a candidate of cryogenic bonding for a mirror material, including sapphire, of the gravitational wave experiments: LIGO, VIRGO, and KAGRA. 
We experimentally studied the mechanical strength of the bonded interface for two different surface conditions: polished and unpolished.
We demonstrated that the tensile and shear strength $>$ 20~MPa for samples with a polished surface. 
This satisfied the requirement of 5.5~MPa derived from the mechanical simulation assuming a launch load of 30G.
We identified that samples glued on a polished surface exhibit higher strength than unpolished ones by a factor of 2 for tensile and 18 for shear strength. 
We measured the millimeter-wave transmittance between $90$ and $140$~GHz using sapphire plates with a diameter of 50~mm before and after bonding.
We did not find any optical effects caused by the bonded interface within 2\% error in transmittance, which originates from the measurement system. 
\end{abstract}

% Include a list of up to six keywords after the abstract

{\noindent \footnotesize\textbf{*}Yuki Sakurai,  \linkable{ysakurai@s.okayama-u.ac.jp} }
% Include email contact information for corresponding author
%{\noindent \footnotesize\textbf{*}Fourth author name,  \linkable{myemail@university.edu} }

\begin{spacing}{2}   % use double spacing for rest of manuscript

\section{Introduction}
\label{sec:intro}  % \label{} allows reference to this section
The measurement of the cosmic microwave background (CMB) B-mode polarization allows to test the inflation paradigm. 
One of the challenges is to control systematic effects at a large angular-scale, and the community has been employing an instrument to modulate the incident linearly polarized light using a continuously rotating half-wave plate (HWP). 
Several ground-based and balloon-borne telescopes employ this polarimetry technique. \cite{Blastpol,EBEX,PB2_1,PB2_2,Simons,CMBS4_1,CMBS4_2}
In 2019, a post-Planck CMB polarization satellite mission, LiteBIRD, is selected as the JAXA/ISAS second strategic large-class mission. \cite{Hazumi}
Despite no atmosphere in space, the stringent stability requirement in a detection chain within a satellite system drives the need for signal modulation; therefore a continuous rotating HWP is required.

One of the standard millimeter-wave birefringent materials is sapphire.
An achromatic HWP (AHWP) based on five-layer stacked sapphire is design for  the LiteBIRD low-frequency telescope (LFT). \cite{Sakurai}
While the thermal and optical properties of a sapphire-based AHWP are well matched for CMB polarimetry are well matched, it has to survive the mechanical stress from the launch impact and vibration for use in a satellite mission. 
A preliminary study (refer to Sec.~\ref{sec:strength_testing} for more details) indicates that a five-layer stacked HWP has a risk of breakage at launch unless the five layers are glued together and are able to mechanically treated as one disk. 
We performed a finite element method (FEM) simulation assuming a launch load of 30G and the required bond strength against the launch vibration was estimated to be about 5.5~MPa.

A cryogenically compatible glue based on hydroxide catalysis bonding for sapphire interfaces has been previously explored as a part of the instrument development for the gravitational experiments: LIGO, VIRGO, and KAGRA\cite{Ushiba_2021}.
Technique using a sodium silicate solution chemically bonds materials by creating a silicate network or 3D alumino-silicate network at a glue interface.
In their study, a flatness of a surface is required less than $\lambda$/4 peak-to-valley flatness (PV), where PV is a flatness parameter defined in Appendix \ref{sec:roughness}.
This PV is expressed using the wavelength $\lambda$ of the light source of the laser interferometer used to measure surface flatness, where $\lambda$ = 633~nm is a typical visible red light wavelength \cite{Elliffe,patent}.
The strength of using this bonding technique on a sapphire-sapphire interface would have been sufficiently high for use in an AHWP if the surface of a sapphire disk, the typical diameter of 500~mm and the thickness of 5~mm, could achieve the flatness at the sub micron level.
In practice, it is difficult to precisely machine the surface accuracy to this level over a large area. 
Our development comprises three main steps. First, we characterize the tensile and shear strength of the sapphire-sapphire interfaces over a small area, $4\times4$~mm$^2$, with and without the polished surface. 
Second, we shall extend this work to a larger area. This is because it is challenging to achieve high precision surface flatness over the large area, and this can potentially prevent from proper bonding.
Finally, we shall prepare the full-scale prototype AHWP and carry out the mechanical test, which simulates the launch impact and vibration as if it flies.
We also ensure the millimeter-wave optical performance after the mechanical test.

In this paper, we focus on the first step of our development plan. We study the mechanical strength at a bonding interface between sapphire surfaces.
The tensile and shear strength were measured for polished and unpolished surfaces to investigate the difference of bond strength for the two different surface conditions.
In addition, the millimeter-wave transmittance was measured to check any optical effect from the bonding.
The frequency range of the measurement is between 90 and 140~GHz, which is a part of the observation frequency band of LiteBIRD LFT\cite{Sekimoto_2020_SPIE}.

\section{Hydroxide Catalysis Bonding}
\subsection{Mechanism} % of bonding}
We summarize the chemical reaction process for bonding sapphires using a sodium silicate solution.\cite{patent,Pheps,Phelps_2,Kumar}
The bonding between sapphires is created in three main steps: etching by hydration, polymerization, and dehydration. 

A sodium silicate solution is placed on the first surface to be bonded, and then, the second surface is brought into contact. 
Hydroxide ions in the solution react with the surfaces, and the alumina ions are freed into the solution from the surfaces according to
\begin{equation}
\label{eq:chemical_step_1}
\ce{Al(OH)3} + \ce{OH^-} \rightarrow \ce{Al(OH)4^-} \, .
\end{equation}
In this process, the number of hydroxide ions in the solution decreases; i.e., the pH of the solution reduces. 
Once the pH drops below 11,\cite{book1} silicate ions in the solution dissociate as
\begin{equation}
\label{eq:chemical_step_2}
\ce{Si(OH)5^-} \rightarrow \ce{Si(OH)4} + \ce{OH-} \, ,
\end{equation}
and a silicate network is formed as
\begin{equation}
\label{eq:chemical_step_3}
\ce{2Si(OH)4} \rightarrow \ce{(HO)3SiOSi(OH)3} + \ce{H2O} \, .
\end{equation}
In parallel, \ce{Al(OH)4^{-}} ions begin to gain bond formation via the dimerization process as
\begin{equation}
\label{eq:chemical_step_4}
\ce{2Al(OH)4^{-}} \rightarrow \ce{(HO)3^{-}AlOAl(OH)3^{-}} + \ce{H2O}  \, .
\end{equation}
The alumina ions in the solution will react with the silicate network. 
As the Al atoms replace some of the Si atoms, a 3D alumino-silicate network is formed as
\begin{equation}
\label{eq:chemical_step_5}
\ce{Si(OH)4} + \ce{Al(OH)4^{-}} \rightarrow \ce{(HO)2AlOSi(OH)3} + \ce{H2O} + \ce{OH^-} \, .
\end{equation}
The bonding between sapphires using a sodium silicate solution is realized by the connection through this chain structure of the 3D alumino-silicate network. 
In the final step, moisture evaporates from the interface, and the bond strength settles down.
According to the references, 4 weeks are recommended to cure at room temperature.\cite{Pheps,Kim,Elliffe}

\subsection{Sample Preparations}
We prepared two categories of sapphire samples. 
For the mechanical strength tests, we prepared polished and unpolished a-cut sapphire bars. 
For tensile and shear strength tests, the sample sizes are 4$\times$4$\times$15 and 4$\times$4$\times$20~mm$^3$, respectively.
For the optical measurement, we prepared polished c-cut sapphire plates of diameter 50.8~mm and thickness 2~mm.
All the samples were scrubbed using cerium oxide paste with de-ionized (DI) water, followed by cleaning the surfaces with sodium bicarbonate paste. 
This process helps to remove peaks and makes the surface hydrophilic allowing a bonding solution to smoothly spread over the entire surface\cite{Alix,LIGO}. 
The surface was also inspected by microscope to confirm the visual absence of particles that could cause a partial bonding failure, followed by rinsing using dust-proof wipes with DI water and ethanol. 
A sodium silicate solution $(\ce{Na2O}$:~9-10\%,~$\ce{SiO2}$:~28-32\%,~$\ce{H2O}$:~58-63\%) was diluted with DI water to create a 1:6 volumetric ratio of a solution according to the recipe introduced in Haughian et al.,\cite{IOP} and 1.0~$\mu\ell$/cm$^2$ of a solution was used for bonding.
After bonding, all the bonded samples were cured for 24~h at room temperature, followed by high-temperature curing for 5~h at 100~$^\circ$C to reduce the curing time \cite{Kim}.
After high-temperature curing, the area where the two surfaces were successfully bonded was visually inspected for all the polished test samples. 
We quantified the result based on the visual inspection using a "fractional bonded areas". This is  a measure of a fractional area properly bonded with respect to the entire surface area that is intended to bond.

We found that a part of the surface area has a fringe pattern that is interpreted as the Newton ring. We think this is originated from the small air gap, i.e.,  unbonded area between the two sapphire surfaces. 
For the unpolished samples, we could not estimate the fractional bonded area by the visual inspections because of the non-transparent surface.

\section{Mechanical Characterization}
\label{sec:strength_testing}
\subsection{Structural Requirement}
The FEM-based stress analysis has been carried out.
In this simulation, we assume a five-layer AHWP with a diameter of 476~mm and a thickness of 4.85~mm each.
We evaluate the stress, deformation, and resonance frequency under sinusoidal vibration, random vibration, impact, and acoustic excitation, assuming that the plate fixed rigidly at the outer rim in this simulation.
The results indicate the risk of breakage in the individual sapphire plate when the five plates are not glued to each other. 
This risk is particularly relevant to the sapphire surface that is patterned with the array of sub-wavelength grating structures for anti-reflection purpose \cite{Sakurai,ryota_pharos} because such a micro structure can potentially degrade mechanical properties of the sapphire. 
Using a glue between the sapphire surfaces, the five-layer sapphire stack can be treated as one plate. 
As a result, the stress, deformation, and resonance frequency becomes tolerable to the launch condition.
When the 10\% of the surface area between the sapphire plates is glued in each boundary, the maximum shear strength is 5.5~MPa at the edge of the plate. 
Thus, we employ this value as the required shear strength for a design guidance for the AHWP to be intact. 
We admit that this requirement depends on how much we glue over the surface area. 
We have taken a conservative choice as 10\%. 
This is because to achieve high surface accuracy over a large sapphire plate, e.g., 476~mm$\times$4.85~mm, is challenge.
Thus, we have a room to adjust this requirement as we learn more about the feasibility of the sapphire surface accuracy in future studies. 
%plates are glued over the 10\% of its surface 
%In order for the five plates to be intact, we derive the required sheared stress to be more than 5.5.~MPa as the design guidance assuming that a glue is applied over 10\% of the sapphire surface area.

% \begin{figure}[h]
%    \centering
%    \includegraphics[width = 0.3\textwidth]{Figure/AHWP_diagram.png}
%    \caption{The conceptual illustration of five-layer %AHWP.\cite{Kunimoto_spie_demo}}
%    \label{fig:AHWP_sketch}
%\end{figure}

\subsection{Measurements}
\begin{figure}[h]
    \centering
    \includegraphics[width = 0.9\textwidth]{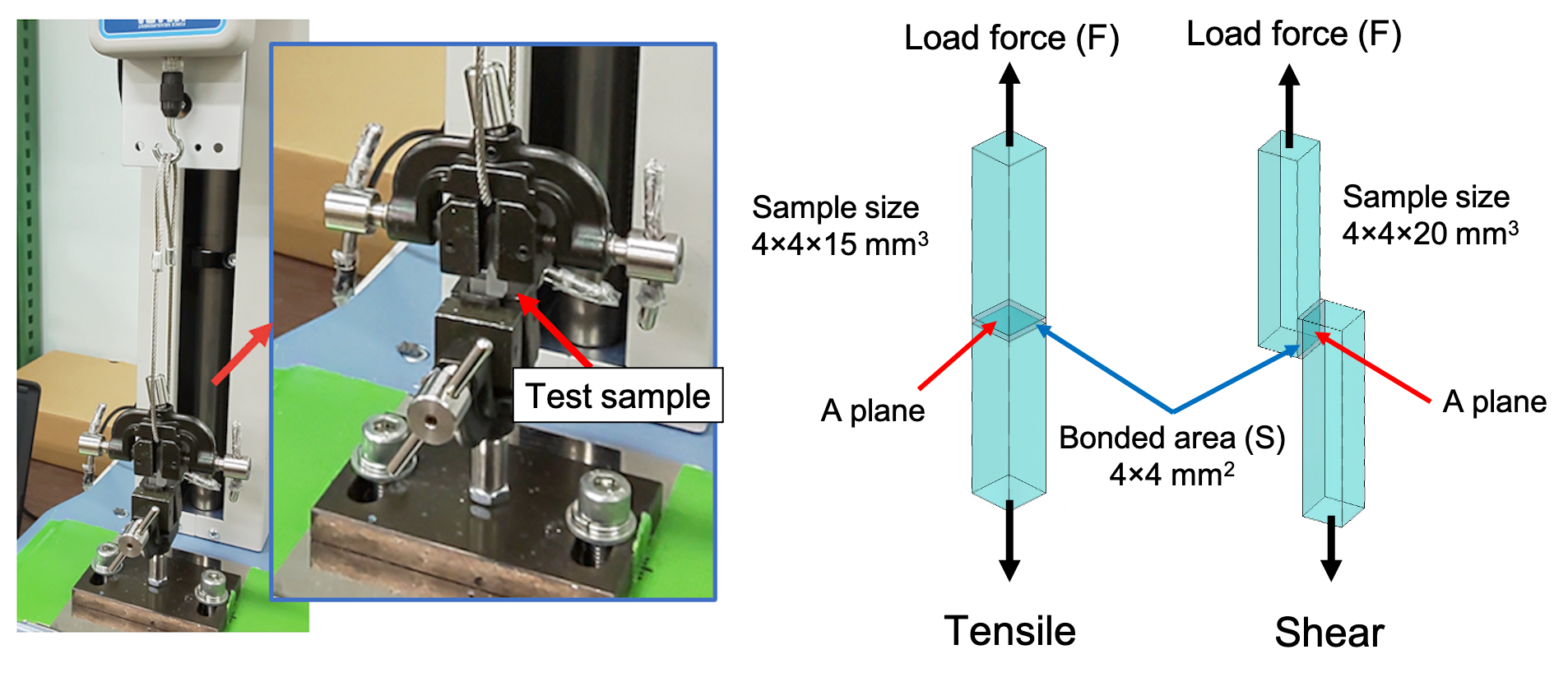}
    \caption{Left: experimental apparatus used in the gluing strength measurement. Right: measurement setup.}
    \label{fig:Strength_testing}
\end{figure}

We carried out two types of the mechanical strength measurement: tensile and shear strength of the two a-cut plane sapphires that are bonded, as shown in the right panel of Figure~\ref{fig:Strength_testing}. 
The sizes of the samples are 4$\times$4$\times$15~mm$^3$ and 4$\times$4$\times$20~mm$^3$ for tensile and shear measurements, respectively. 
For tensile measurements, we glued the a-planes of 4$\times$4~mm$^2$ each.
For shear measurements, we glued the a-plane of 4$\times$20~mm$^2$ with a gluing area of 4$\times$4~mm$^2$.
The polished sample is defined as PV= $\lambda/2 \sim\lambda$ and Ra $\leq$ 1~nm, and the unpolished sample is defined as Ra $\leq$ 1~$\mu$m, where $\lambda$ is a typical visible red wavelength of 633~nm.
See Appendix \ref{sec:roughness} for the definition of PV and Ra.

The test samples are listed in Table~\ref{Table:number_of_samples}. 
Before measuring the mechanical strength, we visually inspected the bonded surface of the polished samples and recorded the fractional bonded area. 
We used a tensile testing machine and a digital force-meter, as shown in the left panel of Figure~\ref{fig:Strength_testing}. 
Force was applied to a bonded sample until it breaks or reaches 450~N, which is limited by the force-meter. 
The maximum loaded force $F_{\rm max}$ was recorded, and the tensile (shear) stress $P$ was calculated by
\begin{equation}
    P = \frac{{F}_{\rm{max}}}{S} \, ,
    \label{eq:1}
\end{equation}
where $S$ is the targeted bonded area of 4$\times$4~mm$^2$.
When the sample does not break at 450~N, the calculated stress using Eq.~\ref{eq:1} is 450~N/16~mm$^2$ $\sim$ 28.1~MPa as the maximum measurable strength.

\subsection{Results}
\begin{figure}[h]
    \begin{center}
    \begin{tabular}{c}
    \includegraphics[width = 0.9\textwidth, trim = 0 0 0 30]{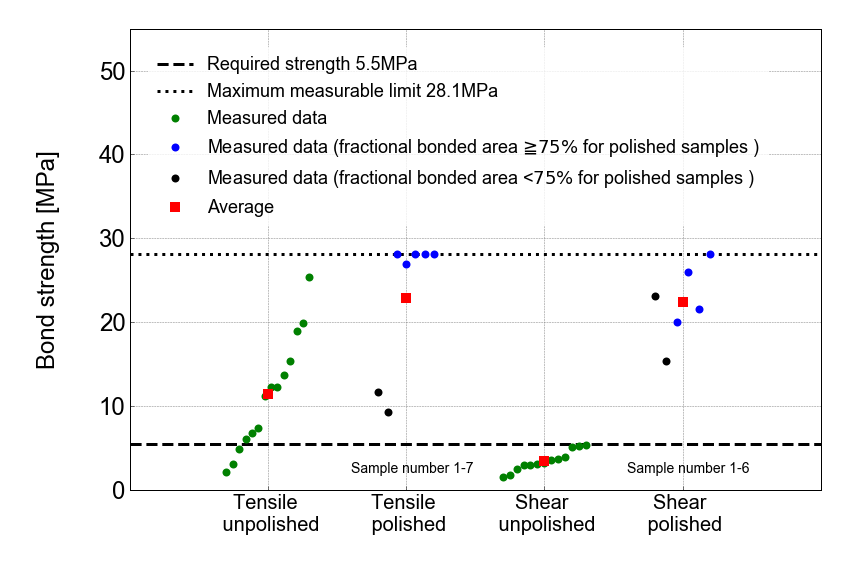}
    \end{tabular}
    \end{center}
    \caption{Results of the tensile and shear strength for polished and unpolished samples. Green points show the measured data of unpolished samples. Blue and black points show the results of polished samples with the fractional bonded area greater and less than 75\%, respectively. Red points indicate the average value in each category.}
    \label{fig:Strength_testing_result_1}
\end{figure}

Figure~\ref{fig:Strength_testing_result_1} shows the tensile and shear strength for the unpolished and polished samples. 
In the polished samples, we highlight the data point that achieves the fractional bonded area of $>$ 75\% as blue data points. 
Red points indicate the average value in each category. 
The bold dashed line shows the required strength of 5.5~MPa as the design guidance. 
The dotted line represents the maximum measurement limit of 28.1~Mpa in this measurement setup.
\begin{table}[h]
    \caption{The number of tested samples for each sample category and polished samples with the fractional bonded area greater than 75\%.}
    \label{Table:number_of_samples}
    \begin{center}
    \begin{tabular}{|c|c|c|c|}
    \hline
    \multicolumn{2}{|c|}{Sample category } & Total \# of tested samples  & 
    \# of samples with bonded area$>$75\%   \\
    \hline\hline
    \multirow{2}{*}{Tensile} & Unpolished & 14 & N.A.\\
    \cline{2-4}
    & Polished & 7 & 5\\
    \hline
    \multirow{2}{*}{Shear} & Unpolished & 13 & N.A.\\
    \cline{2-4}
    & Polished & 6 & 4\\
    \hline
    \end{tabular}
    \end{center}
\end{table}

\begin{table}[h]
    \caption{Results of the fractional bonded area for polished samples and strength results before and after the correction.}
    \label{Table:bond_coverage_strength}
    \begin{center}
    \begin{tabular}{|c|c|c|c|c|}
    \hline
    \rule[-1ex]{0pt}{3.5ex} & Sample number & 
    \begin{tabular}{c}Fractional \\ bonded area (\%) \end{tabular} &
    \begin{tabular}{c}Uncorrected \\ strength (MPa) \end{tabular} &
    \begin{tabular}{c}Corrected \\ strength (MPa) \end{tabular} \\
    \hline\hline
    \rule[-1ex]{0pt}{3.5ex} & No.1 & 35 & 11.7 & 33.4\\
    \cline{2-5}
    \rule[-1ex]{0pt}{3.5ex} & No.2 & 35 & 9.3 & 26.6 \\
    \cline{2-5}
    \rule[-1ex]{0pt}{3.5ex} & No.3 & 98 & 28.1 & 28.7 \\
    \cline{2-5}
    \rule[-1ex]{0pt}{3.5ex} Tensile & No.4 & 98 & 26.9 & 27.4\\
    \cline{2-5}
    \rule[-1ex]{0pt}{3.5ex} & No.5 & 98 & 28.1 & 28.7\\
    \cline{2-5}
    \rule[-1ex]{0pt}{3.5ex} & No.6 & 99 & 28.1 & 28.4\\
    \cline{2-5}
    \rule[-1ex]{0pt}{3.5ex} & No.7 & 95 & 28.1 & 29.6\\
    \hline
    \rule[-1ex]{0pt}{3.5ex}  & No.1 & 70 & 23.1 & 33.0\\
    \cline{2-5}
    \rule[-1ex]{0pt}{3.5ex} & No.2  & 70 & 15.4 & 22.0\\
    \cline{2-5}
    \rule[-1ex]{0pt}{3.5ex} Shear & No.3 & 90 & 20.0 & 22.2\\
    \cline{2-5}
    \rule[-1ex]{0pt}{3.5ex} & No.4 & 85 & 26.0 & 30.6\\
    \cline{2-5}
    \rule[-1ex]{0pt}{3.5ex} & No.5 & 85 & 21.6 & 25.4\\
    \cline{2-5}
    \rule[-1ex]{0pt}{3.5ex} & No.6 & 75 & 28.1 & 37.5\\
    \hline
    \end{tabular}
    \end{center}
\end{table}

\begin{figure}[h]
    \begin{center}
    \begin{tabular}{c}
    \includegraphics[width = 0.9\textwidth, trim = 0 0 0 15]{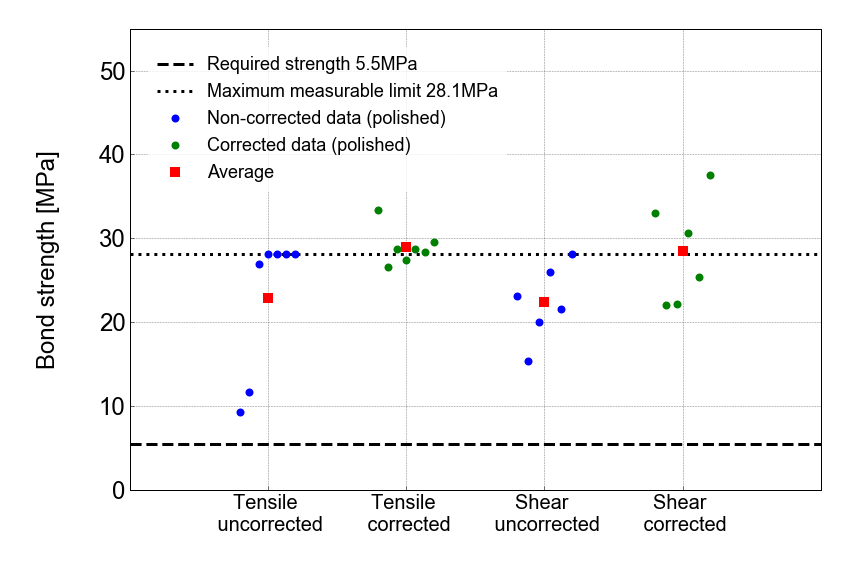}
    \end{tabular}
    \end{center}
    \caption{Comparison of the stress computed with and without the correction from the fractional bonded area. Blue and green data points show the bond strength results for polished samples without and with the correction, respectively. The averaged strength for each sample category is shown as red points.}
    \label{fig:Strength_testing_result_2}
\end{figure}

The tensile strength data for the unpolished samples show the widest spread.
This indicates that a part of unpolished samples are only partially glued or incomplete state of glue, but we do not have a method to confirm. 
We observed three data points are below the required strength.
While the number of data points is limited, the tensile strength data for the polished samples exhibit a tendency of higher strength with a higher fractional bonded area.
The shear strength data for the unpolished samples are consistently lower than the requirement. 
On the other hand, the shear strength data for the polished samples are above the required strength at all data points. 
In Figure~\ref{fig:Strength_testing_result_1}, we computed the stress using $S$ of 4$\times$4~mm$^2$ in Eq.~\ref{eq:1}. 
We hypothesized the cause of the variance in the polished sample data as some of the samples may not be fully bonded over the entire surface of 4$\times$4~mm$^2$. 
Therefore, we estimated the fractional bonded area using visual inspection (Table~\ref{Table:bond_coverage_strength}).
To account for this effect, we replace $S$ in Eq.~\ref{eq:1} with the area that is multiplied with the fractional bonded area. 
The results are listed in Table~\ref{Table:results_table} (Figure~\ref{fig:Strength_testing_result_2}). 
Only the polished sample can estimate the fractional bonded area; thus, no correction has been applied to the unpolished sample data.
\begin{table}[h]
    \caption{Results of the averaged, maximum and minimum bond strength for each sample category.}
    \label{Table:results_table}
    \begin{center}
    \begin{tabular}{|c|c|c|c|c|}
    \hline
    \rule[-1ex]{0pt}{3.5ex}& Sample category & Ave. (MPa) & Max. (MPa) & Min. (MPa)\\
    \hline\hline
    \rule[-1ex]{0pt}{3.5ex}Uncorrected & Tensile (unpolished) & 11.4$\pm$1.8 & 25.4 & 2.1\\
    \cline{2-5}
    \rule[-1ex]{0pt}{3.5ex}& Tensile (polished) & 23$\pm$3 & 28.1 & 9.3\\
    \cline{2-5}
    \rule[-1ex]{0pt}{3.5ex}& Shear (unpolished) & 3.4$\pm$0.3 & 5.3 & 1.5\\
    \cline{2-5}
    \rule[-1ex]{0pt}{3.5ex}&Shear (polished) & 22.4$\pm$1.7 & 28.1 & 15.4\\
    \hline
    \rule[-1ex]{0pt}{3.5ex}Corrected & Tensile (polished) & 29.0$\pm$0.2 & 33.4 & 26.6\\
    \cline{2-5}
    \rule[-1ex]{0pt}{3.5ex}&Shear (polished) & 28.5$\pm$1.1 & 37.5 & 22.0\\
    \hline
    \end{tabular}
    \end{center}
\end{table}

While the variation of the stress over the samples is decreased after the correction using the fractional bonded area, we still observe the variation of data points.
This may be due to the imperfection of the sample alignment with respect to the direction of the force applied to the sample. We have conducted stress measurements by aligning the samples and applied the force with the best effort basis.
The current variation is sufficiently small enough to conclude that the tensile and shear strength for polished samples are well above the required strength.

\subsection{Cryogenic Robustness}

A serious concern of using a glue at cryogenic temperature is a delamination of the glued layers due to the differential thermal contraction.
We prepare two sapphire plates with a diameter of 50~mm each that are bonded and submerged in liquid nitrogen.
This sample was used for the millimeter wave transmittance characterization, which is detailed in Section~\ref{sec:trans}. 
We repeated the following sequence for ten times: 
1) submerge the sapphire disk in the liquid nitrogen, 
2) keep until the boiling subsides, 
3) remove the sample from the liquid nitrogen and quickly warm it to room temperature using a heat gun. 
Despite the rapid cycle of cooling and warming, we did not observe any difference before and after the tests including any delamination nor crack.
This result indicates the robustness of this bonding technique at least with this diameter of two-layered sapphire samples. 
We shall investigate this for a larger sample operating at $<$ 20~K. 

\section{Millimeter Wave Optical Characterization}
\label{sec:trans}
\subsection{Measurements}
\begin{figure}[h]
    \centering
    \includegraphics[width = 0.33\textwidth]{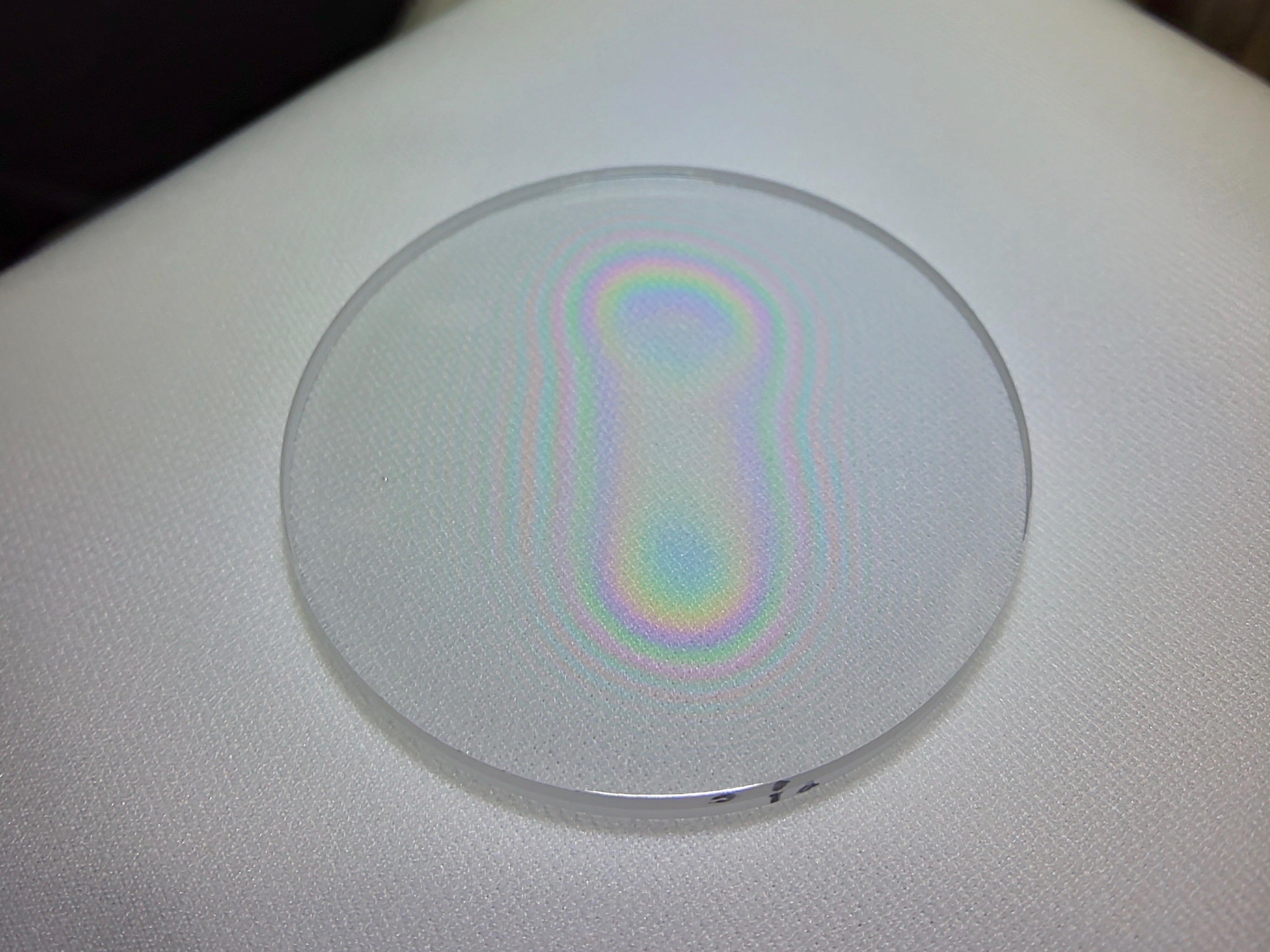}
    \includegraphics[width = 0.33\textwidth]{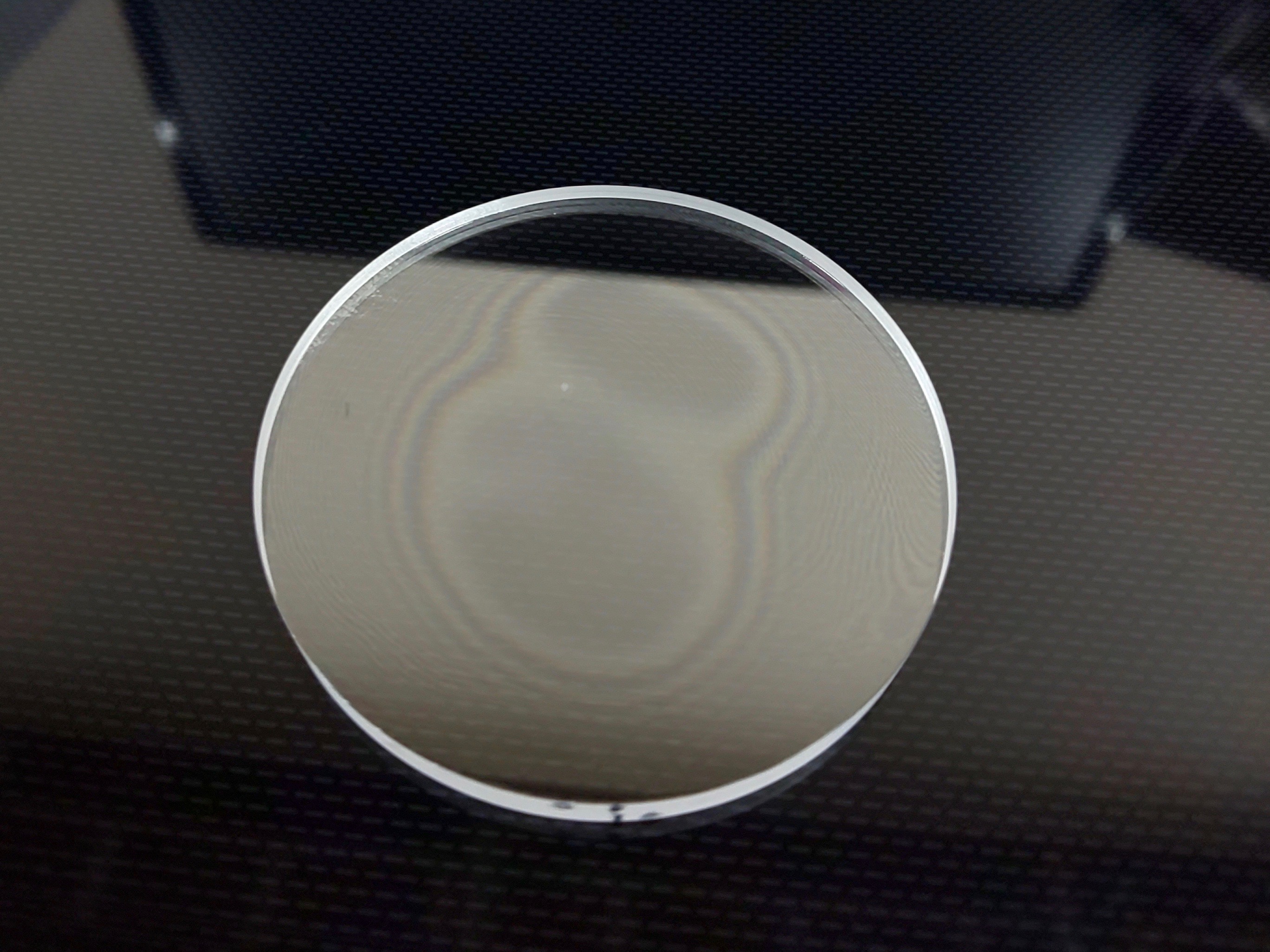}
    \caption{Left: the two c-cut sapphire samples are stacked on top of each other without bonding. Right: the two samples are stacked with bonding.}
    \label{fig:csapphire}
\end{figure}

\begin{figure}[h]
    \centering
    \includegraphics[width = 0.73\textwidth]{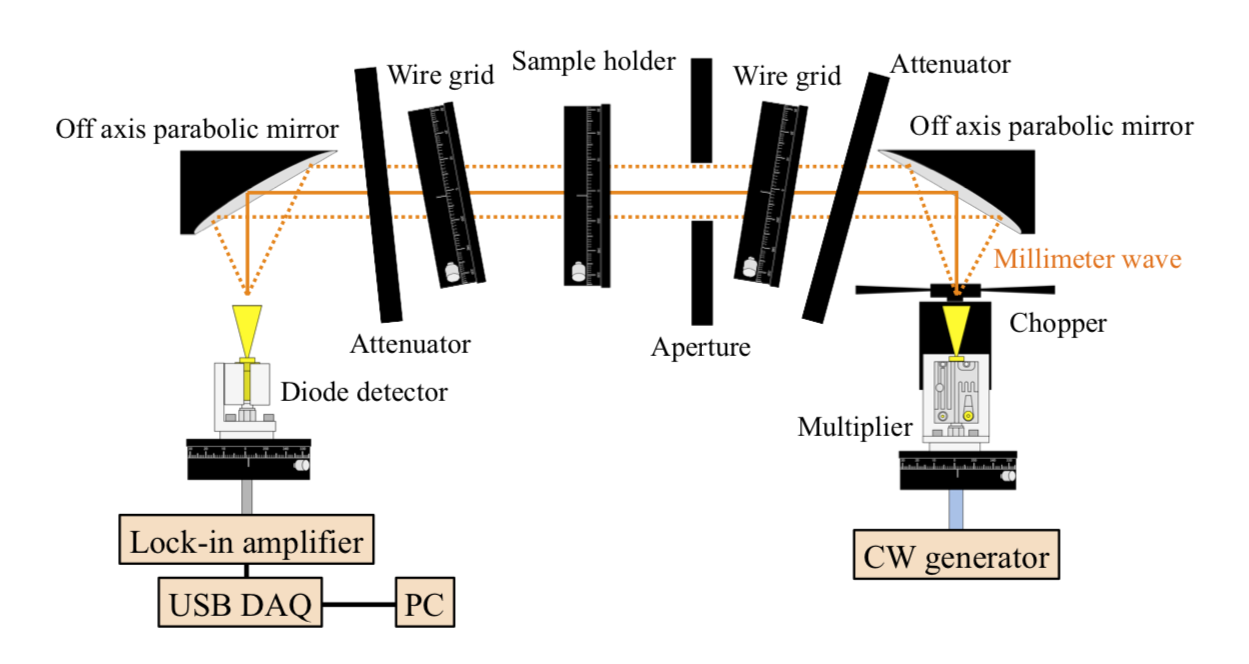}
    \caption{The schematic diagram of the optical measurement system\cite{Komatsu_quote}.}
    \label{fig:optical_system}
\end{figure}

The millimeter wave transmittance of two-layer stacked sapphire plates were measured between 90 and 140~GHz before and after the bonding. 
In this measurement, c-cut sapphire plates, as opposed to a-cut, were used to determine the optical effects caused by the bonding layer.
Two-layered bonded samples with a diameter of 50.8~mm and thickness of 2~mm are shown in Figure~\ref{fig:csapphire}.
The optical measurement system is shown in Figure~\ref{fig:optical_system}.
Further detailed descriptions of the measurement setup have been found by Komatsu et al.\cite{Komatsu_quote}\\


The transmittance at each frequency is the ratio of the power detected with a sample to the power detected without a sample. 
The power at each frequency point is measured as the average of two detector positions separated by $\lambda$/4 along the light pass, where $\lambda$ is the wavelength of input radiation. In this way, we minimize the undesired standing wave present in the measurement system.

Transmittance measurements possess systematic errors due to several sources: the incomplete cancellation of standing waves, the position misalignment of the optical components, the input power instability, and the detector non-linearity.
We estimate the combination of these systematic errors by computing the RMS of residuals obtained by fitting to the transmittance data of a known sample with a theoretical model.
As for the reference, the estimated magnitude of residuals in this measurement setup between 90 and 140~GHz is 2\% provided by Takaku et al.\cite{ryota_pharos}.
Considering the impact of the bonding, we define the fractional difference of measured transmittance with and without a bonding layer between the two plates as
\begin{equation}
    D = (T_2 - T_1)/T_1 \, ,
    \label{eq:3}
\end{equation}
where $T_1$ and $T_2$ are the measured transmittances before and after bonding, respectively.

\subsection{Results}

Transmittance measurements are conducted at three different rotational angles of the sample (0$^\circ$, 120$^\circ$, and 240$^\circ$).
The 0$^\circ$ point is defined as arbitrary, but it is consistent between before and after bonding.
We made a fiducial mark on the side surface of the sapphire disk.
Thus, the three angles before and after bonding are the same with an accuracy of $\pm$ 1$^\circ$.
This is to check any potential rotational-dependence effects, e.g., birefringent effects caused by the presence of glue, an optical axis tilt with respect to sample plane, and so on.

The measured transmittance data are fitted with the followings, the refractive index ($n$), loss tangent ($\tan{\delta}$) and thickness ($d$).
The thickness is varied within the measured error, which is the maximum and the minimum thicknesses measured by a micrometer.
The measured thicknesses of the two samples are $4.034\pm0.010$~mm and $4.033\pm0.009$~mm before and after bonding, respectively.
We observed consistent thickness before and after bonding within the measurement error.

As shown in Figure~\ref{fig:csapphire}, we observed interference fringes.
At minumum of 20 red fringes were observed along the vertical direction from the disk.
We can estimate the order of magnitude of the air gap $d$ between the two sapphire plates if we apply the Newton ring formula as
\begin{eqnarray}
d = \frac{\lambda_r R}{2 \Delta R},
\end{eqnarray}
where $\lambda_r = 633$~nm (visible red light), $R$ is the radius of the sapphire disk, and $\Delta R$ is the distance from one ring to the adjacent ring. 
The typical ring width can span from 1 to 10~mm over the radius, which corresponds to the maximum air gap to be 8~$\mu$m or less. 
This is consistent with the measured non-uniformity of the disk flatness quoted as 9 to 10~$\mu$m. 

\begin{figure}[htbp]
    \centering
    \includegraphics[width = 0.8\textwidth]{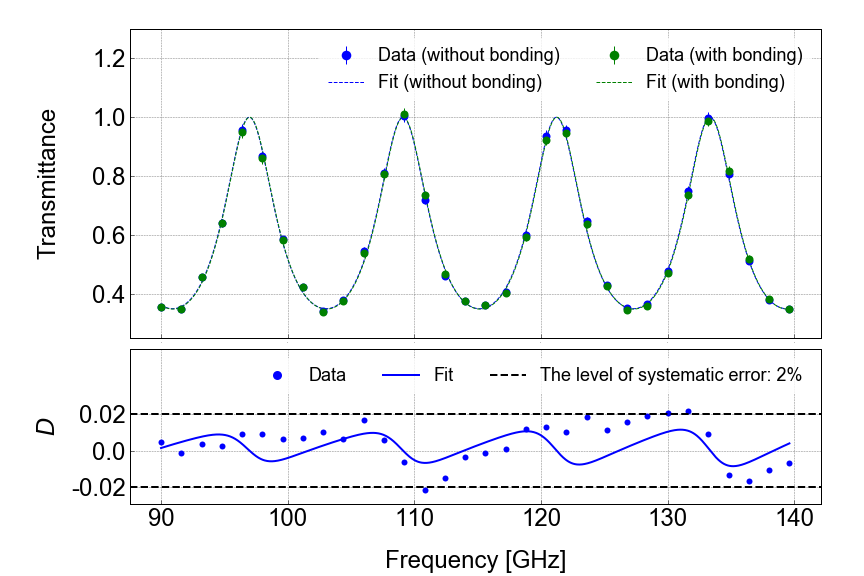}
    \caption{Top: the transmittance spectra of two-layer stacked sapphire plates without (blue) and with (green) bonding at the rotational angles of 0$^\circ$. The measured data and the fit are shown as dot points and dashed lines, respectively. Bottom: the fractional difference between measured data and the fit (blue) with the 2\% systematic error (black dashed line).}
    \label{fig:Transmittance_comp_0deg}
\end{figure}

\begin{figure}[htbp]
    \centering
    \includegraphics[width = 0.8\textwidth]{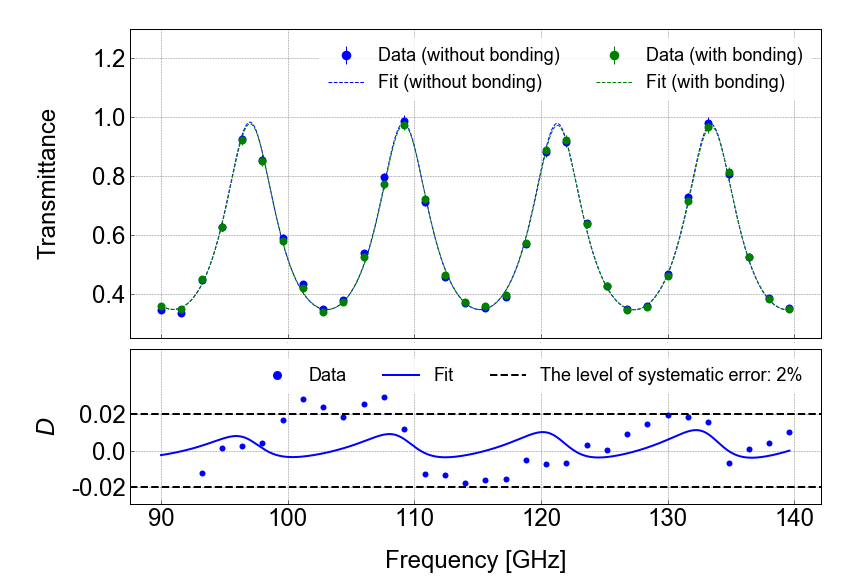}
    \caption{Top: the transmittance spectra of two-layer stacked sapphire plates without (blue) and with (green) bonding at the rotational angles of 120$^\circ$. The measured data and the fit are shown as dot points and dashed lines, respectively. Bottom: the fractional difference between measured data and the fit (blue) with the 2\% systematic error (black dashed line).}
    \label{fig:Transmittance_comp_120deg}
\end{figure}

\begin{figure}[htbp]
    \centering
    \includegraphics[width = 0.8\textwidth]{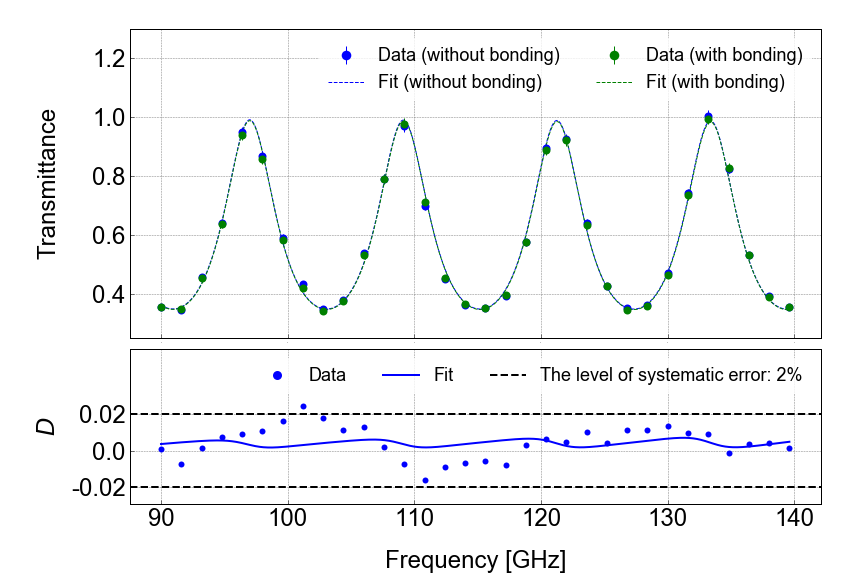}
    \caption{Top: the transmittance spectra of two-layer stacked sapphire plates without (blue) and with (green) bonding at the rotational angles of 240$^\circ$. The measured data and the fit are shown as dot points and dashed lines, respectively. Bottom: the fractional difference between measured data and the fit (blue) with the 2\% systematic error (black dashed line).}
    \label{fig:Transmittance_comp_240deg}
\end{figure}

We fit the data assuming that the two glued sapphires as one plate.
Figure~\ref{fig:Transmittance_comp_0deg},  \ref{fig:Transmittance_comp_120deg} and  \ref{fig:Transmittance_comp_240deg} show the transmittance of the two-layer sapphire with and without the bonding for the rotational angles of 0$^\circ$, 120$^\circ$, and 240$^\circ$, respectively.
The fractional difference $D$ between the measured and fitting data, as shown in the bottom panels of Figure~\ref{fig:Transmittance_comp_0deg},  \ref{fig:Transmittance_comp_120deg} and  \ref{fig:Transmittance_comp_240deg}, are less than the systematic error of 2\%. Therefore, we observed no significant difference before and after bonding. 
The fitting parameters are listed in Table~\ref{Table:fit_parameter}. 
In addition, with regard to the refractive index and the loss tangent, there is no impact caused by bonding within the error.
Here, fitting errors are calculated by assuming that each measured data point has a systematic error of 2\%.

\begin{table}[htbp]
    \caption{Fitted parameters of the refractive index $n$ and loss tangent tan$\delta$ and their fit errors for two cases: without and with bonding. The fit errors are calculated with 2\% error on each data points.}
    \label{Table:fit_parameter}
    \begin{center}
    \begin{tabular}{|c|c|c|c|}
    \hline
    \rule[-1ex]{0pt}{3.5ex}& Rotational angle & $n$ & tan$\delta$($\times{10}^{-4}$)\\
    \hline\hline
    \rule[-1ex]{0pt}{3.5ex} & 0$^\circ$ & 3.058$\pm$0.064 & 0.0$\pm$7.9\\
    \cline{2-4}
    \rule[-1ex]{0pt}{3.5ex}Without bonding & 120$^\circ$ & 3.062$\pm$0.055 & 4.1$\pm$6.8\\
    \cline{2-4}
    \rule[-1ex]{0pt}{3.5ex}& 240$^\circ$ & 3.056$\pm$0.053 & 2.2$\pm$6.4\\
    \hline
    \rule[-1ex]{0pt}{3.5ex}&0$^\circ$ & 3.064$\pm$0.072 & 0.0$\pm$8.9\\
    \cline{2-4}
    \rule[-1ex]{0pt}{3.5ex}With bonding & 120$^\circ$ & 3.056$\pm$0.074 & 4.9$\pm$9.0\\
    \cline{2-4}
    \rule[-1ex]{0pt}{3.5ex}& 240$^\circ$ & 3.063$\pm$0.060 & 2.8$\pm$7.4\\
    \hline
    \end{tabular}
    \end{center}
\end{table}

\section{Conclusions}
We have explored the possibility of using sodium silicate solution to bond surfaces of sapphire for multi layer stacked AHWP. 
We performed mechanical strength and optical test using polished and unpolished sapphire samples. 
The results of the mechanical tests indicated that the tensile and shear strength with a polished surface are higher than 20~MPa.
It is more than the targeted strength of 5.5~MPa, which is derived from the preliminary simulation.
On the other hand, the unpolished samples consistently exhibited lower strength than the polished samples. 
We measured millimeter-wave transmittance to evaluate the effect of the bonding and did not observe the difference in transmittance within the measurement error of 2\% when the two-layer stacked sapphire plates were pre-bonded and post-bonded.
We did not find any result that prohibited us from using hydroxide catalysis bonding techniques for sapphire AHWP to overcome the launch impact.
However, we plan to carry out further the development using the final AHWP size, $\sim$500~mm. The evaluation items are to bond the sapphire plates with imperfect surface accuracy, to check the cryogenic robustness from the  differential thermal contraction, and to test the mechanical tolerance from the launch impact and vibration, and the transmission and polarization performance at the millimeter wavelength. These tests will be carried out in future. 

\subsection* {Acknowledgments}
This study is based on a SPIE conference proceedings \cite{Takayuki_SPIE}. It was supported by JSPS KAKENHI under Grant Nos. JP17H01125, JP18KK0083, JP18J20148,	JP19K14732 and JSPS Core-to-Core Program, A. Advanced Research Networks (Grant No.: JPJSCCA20200003). 
We acknowledge the World Premier International Research Center Initiative (WPI), MEXT, Japan, for support through Kavli IPMU. 
We thank Jun Nakagawa and Koji Kusama for valuable discussions and suggestions.

%%%%% References %%%%%

\bibliography{report}   % bibliography data in report.bib

\begin{thebibliography}{10}

\bibitem{Blastpol}
B.~J. Dober {\em et~al.}, ``{The next-generation BLASTPol experiment},'' in
  {\em Millimeter, Submillimeter, and Far-Infrared Detectors and
  Instrumentation for Astronomy VII},  {\em Proc. SPIE} {\bf 9153}, 137--148
  (2014).

\bibitem{EBEX}
A.~M. Aboobaker {\em et~al.}, ``{The EBEX balloon-borne experiment—optics,
  receiver, and polarimetry},'' {\em The Astrophysical Journal Supplement
  Series} {\bf 239(1)}, 7  (2018).

\bibitem{PB2_1}
D.~Kaneko {\em et~al.}, ``{Deployment of polarbear-2a},'' {\em Journal of Low
  Temperature Physics} {\bf 199(3)}, 1137--1147  (2020).

\bibitem{PB2_2}
C.~A. Hill {\em et~al.}, ``A large-diameter cryogenic rotation stage for
  half-wave plate polarization modulation on the polarbear-2 experiment,'' {\em
  Journal of Low Temperature Physics} {\bf 193}, 851--859  (2018).

\bibitem{Simons}
N.~Galitzki {\em et~al.}, ``{The Simons Observatory: instrument overview},'' in
  {\em Millimeter, Submillimeter, and Far-Infrared Detectors and
  Instrumentation for Astronomy IX},  J.~Zmuidzinas and J.~R. Gao, Eds., {\em
  Proc. SPIE} {\bf 10708}, 1--13  (2018).

\bibitem{CMBS4_1}
K.~N. Abazajian {\em et~al.}, ``{CMB-S4 Science Book, First edition},''
  (2016).

\bibitem{CMBS4_2}
M.~H. Abitbol {\em et~al.}, ``{CMB-S4 Technology Book, First edition},''
  (2017).

\bibitem{Hazumi}
M.~Hazumi {\em et~al.}, ``{LiteBIRD satellite: JAXA's new strategic L-class
  mission for all-sky surveys of cosmic microwave background polarization},''
  in {\em Space Telescopes and Instrumentation 2020: Optical, Infrared, and
  Millimeter Wave},  {\em Proc. SPIE} {\bf 11443}, 114432F  (2020).

\bibitem{Sakurai}
Y.~Sakurai {\em et~al.}, ``{Breadbord model of the polarization modulator unit
  based on a continuous rotating half-wave plate for the low-frequency
  telescope of the LiteBIRD space mission},'' in {\em Millimeter,
  Submillimeter, and Far-Infrared Detectors and Instrumentation for Astronomy
  X},  {\em Proc. SPIE} {\bf 11453}, 114534E  (2020).

\bibitem{Ushiba_2021}
T.~Ushiba {\em et~al.}, ``{Cryogenic suspension design for a kilometer-scale
  gravitational-wave detector},'' {\em Classical and Quantum Gravity} {\bf 38},
  085013  (2021).

\bibitem{Elliffe}
E.~J. Elliffe {\em et~al.}, ``Hydroxide-catalysis bonding for stable optical
  systems for space,'' {\em Class. Quantum Grav.} {\bf 22}, S257--S267  (2005).

\bibitem{patent}
D.-H. Gwo, ``Ultra precision and reliable bonding method,''  (2001).
\newblock US Patent 6,284,085.

\bibitem{Sekimoto_2020_SPIE}
Y.~Sekimoto {\em et~al.}, ``{Concept design of low frequency telescope for CMB
  B-mode polarization satellite LiteBIRD},'' in {\em Millimeter, Submillimeter,
  and Far-Infrared Detectors and Instrumentation for Astronomy X},   {\bf
  11453}, 189 -- 209, SPIE  (2020).

\bibitem{Pheps}
M.~Phelps {\em et~al.}, ``The strength of hydroxide catalysis bonds between
  sapphire, silicon, and fused silica as a function of time,'' {\em Phys. Rev.
  D} {\bf 98(12)}, 122003  (2018).

\bibitem{Phelps_2}
M.~Phelps {\em et~al.}, ``Strength of hydroxide catalysis bonds between
  sapphire, silicon, and fused silica as a function of time,'' {\em Phys. Rev.
  D} {\bf 98}, 122003  (2018).

\bibitem{Kumar}
R.~Kumar {\em et~al.}, ``Status of the cryogenic payload system for the {KAGRA}
  detector,'' {\em J. Phys. Conf. Ser.} {\bf 716}, 012017  (2016).

\bibitem{book1}
R.~Iler, {\em {{The Chemistry of Silica}}}, Wiley-Interscience Publication
  (1979).

\bibitem{Kim}
H.~S. Kim and T.~L. Schmitz, ``Shear strength evaluation of hydroxide catalysis
  bonds for glass-glass and glass-aluminum assemblies,'' {\em Precis. Eng.}
  {\bf 37}, 23--32  (2013).

\bibitem{Alix}
A.~Preston, B.~Balaban, and G.~Mueller, ``{Hydroxide-bonding strength
  measurements for space-based optical missions},'' {\em Int. J. Appl. Ceram.
  Technol.} {\bf 5(4)}, 365--372  (2008).

\bibitem{LIGO}
H.~Armandula {\em et~al.}, ``{LIGO-E050228-v2: Silicate bonding procedure}.''
  LIGO document, 31 August 2010.

\bibitem{IOP}
K.~Haughian {\em et~al.}, ``The effect of crystal orientation on the cryogenic
  strength of hydroxide catalysis bonded sapphire,'' {\em Class. Quantum Grav.}
  {\bf 32}, 075013  (2015).

\bibitem{ryota_pharos}
R.~Takaku {\em et~al.}, ``{Broadband, millimeter-wave anti-reflective
  structures on sapphire ablated with femto-second laser},'' {\em J. Appl.
  Phys.} {\bf 128(22)}, 225302  (2020).

\bibitem{Komatsu_quote}
K.~Komatsu {\em et~al.}, ``{Demonstration of the broadband half-wave plate
  using the nine-layer sapphire for the CMB polarization experiment},'' {\em
  Journal of Astronomical Telescopes, Instruments, and Systems} {\bf 5(4)},
  1--14  (2019).

\bibitem{Takayuki_SPIE}
T.~Toda {\em et~al.}, ``{Mechanical strength and millimeter-wave transmission
  spectrum of stacked sapphire plates bonded by sodium silicate solution},'' in
  {\em Millimeter, Submillimeter, and Far-Infrared Detectors and
  Instrumentation for Astronomy X},  {\em Proc. SPIE} {\bf 11453}, 788--799,
  SPIE  (2020).

\end{thebibliography}
\bibliographystyle{spiejour}   % makes bibtex use spiejour.bst

\appendix
\section{Definition of roughness parameters}
\label{sec:roughness}
\begin{figure}[h]
    \centering
    \includegraphics[width = \textwidth]{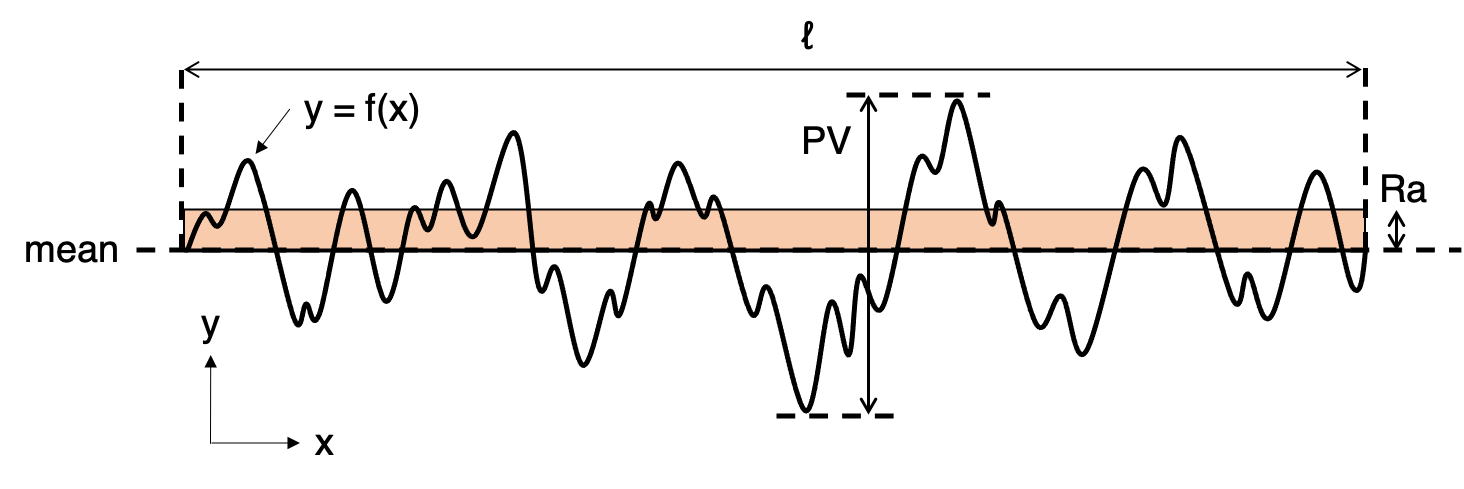}
    \caption{Surface roughness}
    \label{fig:diagram_roughness}
\end{figure}

We represent the surface roughness using two parameters: Peak to Valley (PV) and Ra.
As shown in Figure \ref{fig:diagram_roughness}, the PV is defined as the distance in height between the highest and lowest points on the surface.
In general, the PV is expressed like $\lambda/4$ or $\lambda/20$, where  $\lambda$ is the wavelength of the light source of the laser interferometer used to measure surface accuracy.
$\lambda=633$~nm is a  typical visible red light.

Ra is described as 
\begin{equation}
Ra = \frac{1}{\ell}\int^{\ell}_{0} |f(x)| dx,
\end{equation}
where $\ell$ is the length of the range that defines the surface roughness and $f(x)$ is the function of the roughness curve.
This parameter represents the average roughness over a range.

\section{Transmittance fit model}
\label{sec:fitmodel}

\begin{figure}[h]
    \centering
    \includegraphics[width = \textwidth]{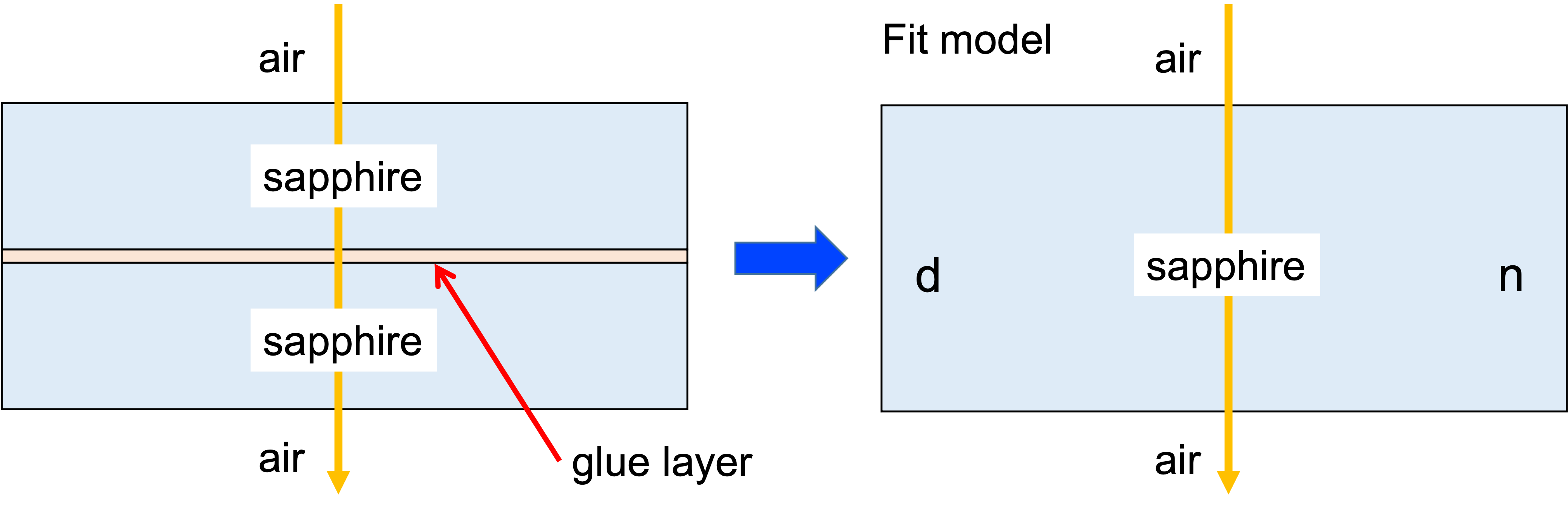}
    \caption{Left: the diagram of the sample with two sapphire layer and gluing layer. Right: the diagram of the fit model where $d$ is the thickness and $n$ is the reflactive index.}
    \label{fig:diagram_trans}
\end{figure}

The measured transmittance is fitted by the theoretical model with free parameters of the refractive index ($n$) and the loss-tangent (tan$\delta$). 
The sample has two sapphire layers before bonding, and an additional layer is added after gluing between them (Figure~\ref{fig:diagram_trans}).
We used a model assuming them as one layer.
We compared the fit results of before and after bonding to check the effect of gluing layer. 

The model equation for the transmittance is 
\begin{equation}
    T = |\frac{2{\gamma}_0}{{{\gamma}_0}{m}_{11} + {\gamma}_0^2{m}_{12} + {m}_{21} + {\gamma}_0{m}_{22}}|^2 \, ,
    \label{eq:transeqs}
\end{equation}
\begin{equation}
    {\gamma}_0 = \sqrt{\frac{{\varepsilon}_0}{{\mu}_0}} \, ,
\end{equation}
Each component in Eq.~\ref{eq:transeqs} is given as
\begin{equation}
    \left(
    \begin{array}{cc}
      {m}_{j,11} & {m}_{j,12}\\
      {m}_{j,21} & {m}_{j,22}\\
    \end{array}
    \right)
    =
    \left(
    \begin{array}{cc}
      \cos(knd) & i\sin(\frac{knd}{{\gamma}})\\
      i{\gamma}\sin(knd) & \cos(knd)\\
    \end{array}
    \right) \, ,
\end{equation}

\begin{equation}
    {\gamma} = {\gamma}_{0}n'
    \simeq {\gamma}_{0}n\left(1-i\frac{1}{2}\tan\delta\right) \, ,   
\end{equation}

where ${k}$ is the wave-number, $n'$ is the complex refractive index, $\tan\delta$ is the loss tangent, and $d$ is the thickness of a layer.
The complex refractive index is approximated using the refractive index and the loss tangent because the loss tangent of sapphire is negligible.

%%%%% Biographies of authors %%%%%

%%\vspace{2ex}\noindent\textbf{First Author} is an assistant professor at the University of Optical Engineering. He received his BS and MS degrees in physics from the University of Optics in 1985 and 1987, respectively, and his PhD degree in optics from the Institute of Technology in 1991.  He is the author of more than 50 journal papers and has written three book chapters. His current research interests include optical interconnects, holography, and optoelectronic systems. He is a member of SPIE.

%\vspace{1ex}
%\noindent Biographies and photographs of the other authors are not available.

\listoffigures
\listoftables

\end{spacing}
\end{document}